\documentclass[%
 reprint,
 amsmath,amssymb,
 aps,
prb,
]{revtex4-2}

\usepackage{graphicx}
\usepackage{dcolumn}
\usepackage{bm}
\usepackage{braket}
\usepackage{changes}

\usepackage{comment}
\usepackage{hyperref}       
\hypersetup{
    colorlinks=true,
    linkcolor=blue,
    filecolor=magenta,      
    urlcolor=blue,
}
\usepackage{color,soul}

\begin{document}

\preprint{APS/1}

\title{Optical Ranging Using Coherent Kerr Soliton Dual-microcombs with Extended Ambiguity Distance}

\author{Yuechen Yang$^{1}$}%
\author{Yang Shen$^{1}$}%
\author{Kailu Zhou$^{1}$}
\author{Chenhua Hu$^{2}$}
\author{Yuanzhuo Ding$^{1}$}
\author{Tinghao Jiang$^{1}$}
\author{Wei Li$^{3}$}%
\author{Yudong Li$^{3}$}%
\author{Liangsen Feng$^{3}$}%
\author{Tengfei Wu$^{3}$}%
\author{Guangqiang He$^{1,4}$}%
\email{gqhe@sjtu.edu.cn}	
\affiliation{%
 $^1$State Key Laboratory of Advanced Optical Communication Systems and Networks, Department of Electronic Engineering, Shanghai Jiao Tong University, Shanghai 200240, China\\
 $^2$State Key Laboratory of Advanced Optical Communication Systems and Networks, School of Physics and Astronomy, Shanghai Jiao Tong University, Shanghai 200240, China \\
 $^3$Science and Technology on Metrology and Calibration Laboratory, Changcheng Institute of Metrology $\rm \&$ Measurement, Aviation Industry Corporation of China, Beijing 100095, China\\
 $^4$SJTU Pinghu Institute of Intelligent Optoelectronics, Department of Electronic Engineering, Shanghai Jiao Tong University, Shanghai 200240, China
}%
	
\date{\today}


\maketitle

\textbf{Optical ranging is a key technology in metrology. Optical frequency combs are shown to provide several advantages in light ranging, offering high precision with high acquisition rate. However, performance of traditional ranging systems based on microcombs is limited by the short ambiguity distance and non-real-time processing. Here, we show that dual-comb ranging system using coherent Kerr soliton microcombs and optical switch realizes extended ambiguity distance and provides a route to real-time processing. The ambguity distance is extended to 3.28 m from about 1.5 mm and the uncertainty reaches about $1.05\times10^{-7}$, while the system is compatible with low-bandwidth detectors. Combining coherent microcomb ranging systems with special FPGA could enable comb-based real-time ranging systems for several applications such as industrial process monitoring.}

\section{Introduction}
Laser based light detection and ranging (LIDAR) is a key technology in industrial and scientific metrology, providing high-precision, long-range and fast-acquisition measurement\cite{10.1117/1.1330700}. Lidar systems have entered a wide range of applications, including autonomous driving, satellite formation flying, unmanned aerial vehicle navigation or gravitational wave detection. When it comes to fast and precise ranging over long distances, optical frequency combs\cite{OpticalFrequencyMetrology} exhibit significant advantages, utilizing time-of-flight methods\cite{Minoshima:00}, interferometric methods\cite{PMID:27558016}, or their combination schemes\cite{rapid-and-precise}. Early experiments utilized the stability of repetition frequency in mode-locked lasers to achieve time-of-flight ranging\cite{Minoshima:00}. The interferometric dual-comb scheme relies on multi-heterodyne detection through the coherent superposition of a pair of frequency combs with slighty different repetition rate ($f_{rep}$)\cite{Dual-CombRanging}, using phase information of beat notes between multiple longitudinal modes. When interference happens between such combs, in the frequency domain, the spectrum of the optical frequency comb is compressed, while in the time domain, the original short pulse signal is stretched\cite{PhysRevA.82.043817}. 
By controlling the repetition rate difference between two optical frequency combs, the broadband optical comb can be compressed into the RF band to form a `radio frequency comb' (RF comb). The spacing between two comb teeth of the electric frequency comb is equal to the repetition rate difference between the two optical combs, which is also the refresh frequency of the ranging system.

The comb-based coherent laser ranging system offers high precision and fast distance acquisition. The dual-comb schemes based on fiber lasers provides methods to achieve extended ambiguity range, real-time processing and fast acquisition rate\cite{rapid-and-precise,fellinger2021simple,li2022ultra,Lang2023freerunning}. However, such optical frequency comb generators have low integration level. In recent years, disspative Kerr soliton (DKS) frequency combs (microcombs) based on high-Q microresonators shows high integration, low power consumption and higher repetition rate\cite{Temporal-solitons}. Microcombs provides a route to ranging systems that combine accuracy and ultra-high acquisition rate\cite{doi:10.1126/science.aao3924}. DKS is generated in a continuous laser driven microresonator, relying on double balance between dispersion and nonlinearity as well as cavity loss and parametric gain\cite{doi:10.1126/science.aan8083}. Mathematically, it is regared as an equilibrium solution of Lugiato-Lefever equation\cite{LLE}. However, existing dual-comb ranging based on microcombs systems are unable to fully possess these positive features: high-integration level, low requirement to detectors, absolute ranging in long distance, fast acquisition rate and real-time processing. The scheme in \cite{doi:10.1126/science.aao3924} have short ambiguity distance $c/2n f_{rep} \approx 1.5\,mm$ and demand for high bandwidth detectors. Scheme in \cite{doi:10.1126/science.aao1968} achieves low-bandwidth compatibility using coherent microcombs, but the aquisition rate varies in several kHz. An important reason for the limitations is that the microcomb is almost immutable because it is generated in a microresonator with certain cavity length and free spectral range (FSR). In other words, the carrier envelope offset frequency ($f_{CEO}$) and $f_{rep}$ of the microcomb are almost fixed. The most important requirement for the detectors is that the bandwidth should cover the beat notes of two microcombs. The center frequency of beat notes between two free-running microcombs may reach a maximum of 1/2 FSR, which generally varies from several GHz to THz\cite{doi:10.1126/science.aay3676}, which places high demands on the detection devices in the system and makes it impossible to achieve real-time processing. Meanwhile, it is proved that a quantified trade-off between precision and ambiguity distance exists in dual-comb ranging systems with certain repetition rate difference ($\Delta f_{rep}$)\cite{Martin:22}. Though the tunable $\Delta f_{rep}$ in several kHz is achieved by changing the pump power
\cite{doi:10.1126/science.aao1968}, the larger continuously tunable range is needed. Consequently, the limitations of microcomb ranging systems need to be overcome.

Here, we show that dual-comb ranging with extended ambiguity distance and low-cost detectors can be achived through coherent DKS microcombs and optical switch, which has the potential for real-time measurement. The coherent microcombs is realized through thermo-optic effect, which extend the light path with the rising of environment temperature by increasing the refractive index of the medium. Therefore, $f_{rep}$ and central frequency of microcombs can be adjusted by controlling temperature. $f_{rep}$ varies in the range of several MHz. The center frequency of the RF comb can be tuned by generating microcombs using pump laser with certain frequency difference. Moreover, according to vernier effect, the ambiguity distance can be extended to $c/2n\Delta f_{rep}$ by switching the two microcombs, where  $\Delta f_{rep}$ stands for the repetition rate difference between two microcombs\cite{rapid-and-precise}. We detail in section 2 the interferometric method with extended ambiguity distance used to measure the distance. The principles of distance measurement and extending ambiguity distance through vernier effect are given. The necessity of using coherent microcombs to achieve compatibility with low-bandwidth detectors is discussed. In section 3, we detail the method of generating coherent microcombs 
by temperature tuning. The tendency of resonant frequency and tunable range of $f_{rep}$ with varying temperature is measured. Finally, in section 4, coherent microcombs are utilized in the ranging system. An comparison between the performance of using two free-running microcombs and coherent microcombs are illustrated, presenting great advantages of reducing phase noise and requirement of detectors. 

\section{Principles of dual-comb ranging with extended ambiguity distance}

\begin{figure*}[t]
    \centering
    \includegraphics[width=\textwidth]{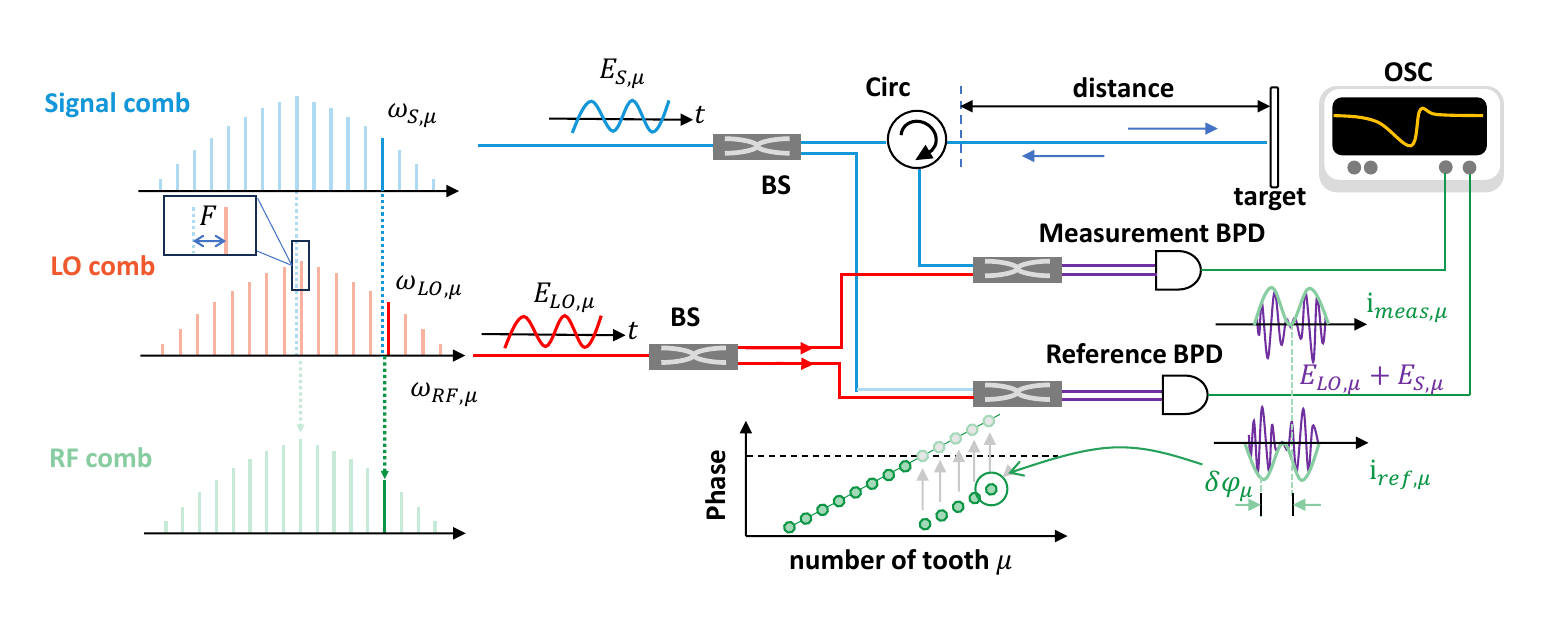}
    \caption{Principle of interferometric dual-comb ranging. The interference between the signal comb and the LO comb generates an RF comb. Each comb tooth of the RF comb is a beat note of one comb tooth of the signal comb and one comb tooth of the local oscillator comb. The interference of the $\mu$-th comb tooth happens in both paths and the measured distance is contained in the phase difference. BS, beam splitter; Circ, circulator; OSC, oscilloscope; BPD, balanced photodetector.}
    \label{principle-of-measurement}
\end{figure*}

We choose the interferometric method for distance measurement. However, in the existed schemes, a critical limitation is the ambiguity range. The basic principle and the method for extending the ambiguity range are discussed in this section. The basic principle of interferometric method is shown in Fig.\ref{principle-of-measurement}, which is based on multi-longitudinal mode heterodyne phase information. The phase-based dual-comb ranging requires two optical frequency combs with slightly different repetition rate, which are used as signal comb and local oscillator (LO) comb, respectively. For the the signal comb, the beam splitter (BS) is used to divide it into two parts, one part is sent to the target to be measured through a circulator (Circ), and then enters the circulator for return after reflection. For ease of expression, the path passing through the measured distance is called `measurement path', while the other path is called `reference path'. For the local oscillator comb, it is directly divided into two parts via a beam splitter. Two signal beams are coupled with two LO beams through couplers, and then enter the balanced photodetector. The signals of the two balanced photodetectors are connected to two channels of the high-speed oscilloscope for recording. In the frequency domain, two microcombs with slightly different $f_{rep}$ enable multi-heterodyne detection. As shown in Fig.\ref{principle-of-measurement}, the optical frequency combs in the THz scale are converted to a RF comb in GHz scale.

To give a brief mathematical description, we select the $\mu$-th spectral line of the signal comb and the local oscillator comb, respectively. The following descriptions are for a single spectral line. For the two parts of the signal comb, one part generates a phase difference $\delta\varphi_{\mu}$ relative to the other part which is determined by the measured distance $d$, the refractive index $n$ of the medium in the measured distance, and the line frequency $\omega_ {s,\mu}.$ The relationship is
\begin{equation}
  \delta\varphi_{\mu} = \frac{2dn(\omega_{s,\mu})\omega_{s,\mu}}{c}
  \label{eq:1}
\end{equation}

The optical field of the measurement path and the reference path satisfies the following equation
\begin{equation}
  E_{S,mes} = E_{S,ref}e^{-j\delta\varphi_{\mu}}
\label{eq:2}
\end{equation}

The two parts of the local oscillation comb coupled to the signal comb are also written as $E_{LO,mes}$ and $E_{LO,ref}$ respectively. Then the beat signal of both paths enter the balanced photodetectors (BPDs). The output signals of the two balanced photodetectors are (detail see \cite{doi:10.1126/science.aao3924})
\begin{equation}
  i_{mes}(t) = \mathcal{R} \frac{A}{Z_0} \mathfrak{T}\left\{E_{S,mes}^* E_{LO,mes}\right\}
\label{eq:3}
\end{equation}
\begin{equation}
      i_{ref}(t) = \mathcal{R} \frac{A}{Z_0} \mathfrak{T}\left\{E_{S,ref}^* E_{LO,ref}\right\}
\label{eq:4}
\end{equation}
where $\mathcal{R}$ is the sensitivity of balanced photodetectors (BPDs), $A$ is their photosensitive area, $Z_0$ is the free-space wave inpedance, and $\mathfrak{T}$ means taking the imaginary part.

After Fourier transform to the signals, the phase difference between the reference and measurement signal is written as
\begin{equation}
  \Delta\Phi = arg\{E_{S,mes}^* E_{LO,mes}\} - arg\{E_{S,ref}^* E_{LO,ref}\}
\label{eq:5}
\end{equation}

Since two parts of the LO comb are identical, $E_{LO,mes}$ and $E_{LO,ref}$ has no phase difference. We have
\begin{equation}
  \Delta\Phi = arg\{E_{S,mes}^*\} - arg\{E_{S,ref}^*\} = \delta\varphi_{\mu}
  \label{eq:6}
\end{equation}

Therefore, we obtained the phase modulation caused by the measurement distance by detecting the signal comb through a multi-heterodyne phase measurement. Combining equation (\ref{eq:1}) and equation (\ref{eq:6}), with $\omega_{s,\mu} = \omega_{s,0}+\mu \omega_{s,r}$, the relationship between the phase difference and the spectral line number is
\begin{equation}
  \Delta\Phi_{\mu} =  \mu\frac{2dn(\omega_{s,\mu})\omega_{s,r}}{c} + \frac{2dn(\omega_{s,\mu})\omega_{s,0}}{c}
\label{eq:7}
\end{equation}
where $\omega_{s,r}$ and $\omega_{s,0}$ correspond to $f_{rep}$ and $f_{\mu = 0}$ of the signal comb, respectively. It can be seen that the phase difference $\Delta\Phi_{\mu}$ is linearly related to the spectral line number $\mu$. In the slope term, except for the distance $d$ to be measured, the refractive index of air, the velocity of light, and $f_{rep}$ of the signal comb are all known. The distance can be calculated by linearly fitting $\Delta\Phi_{\mu}$ to $\mu$.

\begin{figure}[h]
    \centering
    \includegraphics[width=\linewidth]{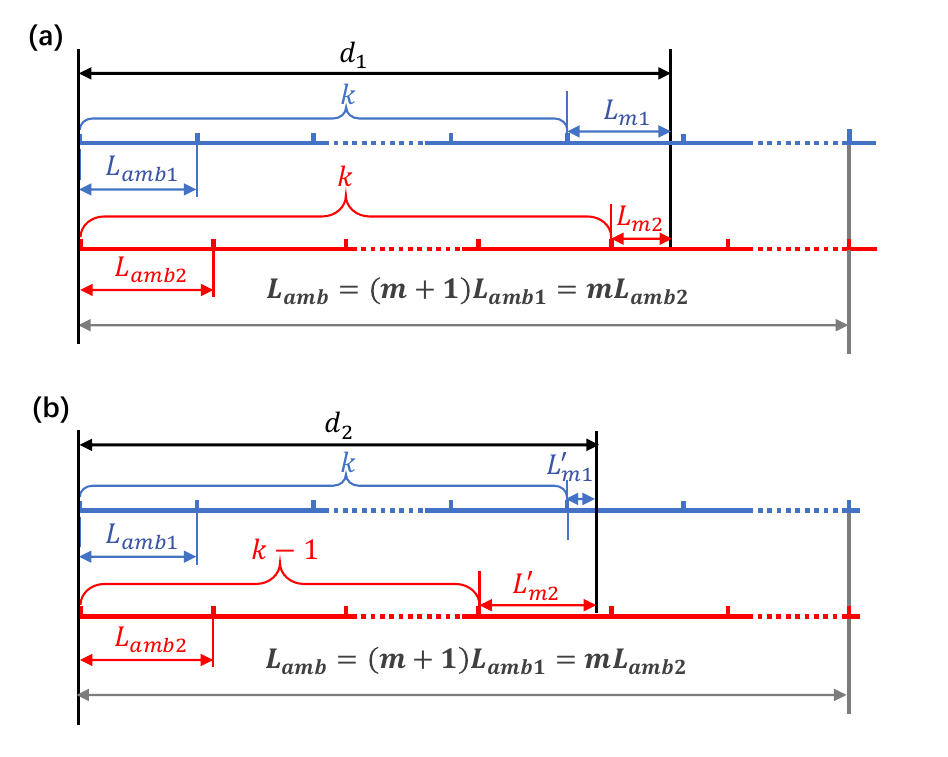}
    \caption{Extend the ambiguity distance via vernier effect. Before and after switching the signal and LO combs, the distance is measured at two different ambiguity distance ($L_{amb1}$ and $L_{amb2}$, $L_{amb2}$ is slightly longer than $L_{amb1}$). Two possible cases during distance synthesis, $L_{m1} > L_{m2}$ and $L_{m1} < L_{m2}$, are shown in (a) and (b), respectively.}
    \label{VernierEffect}
\end{figure}

An optical switch module with 2-input and 2-output is added between the optical frequency comb generation system and the ranging system of the experimental setup. The optical switch has two modes which are directly output and exchange the two optical inputs before outputting. The ambiguity distance of the dual optical comb ranging can be extended through the vernier effect. Without the optical switch, the ambiguity distance is equal to half of the distance between two soliton pulses, which is $L_{amb}=c/2nf_{rep}$. For the microresonator used in our experiment, this distance is approximately 1.5mm, which is difficult to meet the needs of practical applications. We connect the two inputs of the optical switch to soliton frequency comb generation systems, and then connect the output as two optical combs to the subsequent ranging system. By changing the mode of the optical switch, the signal comb and the local oscillator comb are exchanged. In other words, we finish two measurements with the combs. As mentioned before, the two combs have slightly different repetition rates, which cause the ambiguity distance is slightly different in two measurements (note as $L_{amb1}$ and $L_{amb2}$). The principle is shown in Fig.\ref{VernierEffect}, which is similar to multiple pulse repetition frequency method in radar \cite{Modern-radar-system-analysis}. $L_{amb1}$ is slightly shorter than $L_{amb2}$. The $L_{amb}$ 
is extended to the distance between two aligned positions, which is written as $L_{amb} = (m+1) L_{amb1} = m L_{amb2}$. Since the two repetition rates are approximately the same, the expansion factor $m$ can be written as $m = f_{rep}/ \Delta f_{rep}$. The ambiguity distance is calculated by
\begin{equation}
    L_{amb}=  \frac{L_{amb1} L_{amb2}}{L_{amb2}-L_{amb1}}=\frac{c}{2n \Delta f_{rep}}
\label{eq:8}    
\end{equation}
The absolute distance in $L_{amb}$ is calculated through the following rules. First, two ranging results under different ambiguity distances are obtained before and after the switching process. Then, they can be synthesized into one according to vernier effect. Within one ambiguity distance, two different cases may be met: $L_{m1} > L_{m2}$ and $L_{m1} < L_{m2}$. For the former case (shown in Fig.\ref{VernierEffect}(a)), the absolute distance can be regarded as the solution to equation(\ref{L1>L2}).
\begin{equation}
    d_1=kL_{amb1}+L_{m1}=kL_{amb2}+L_{m2}
\label{L1>L2}
\end{equation}
where k is an integer. The measured distance $d_1$ can be derived as
\begin{equation}
  d_1 = \frac{L_{m1}L_{amb2} - L_{m2}L_{amb1}}{L_{amb2} - L_{amb1}}
\label{eq:9}
\end{equation}
For the latter case (shown in Fig.\ref{VernierEffect}(b)), the distance is the solution to equation(\ref{L1<L2}).
\begin{equation}
    d_2=kL_{amb1}+L_{m1}^{'}=(k-1)L_{amb2}+L_{m2}^{'}
\label{L1<L2}
\end{equation}
The measured distance $d_2$ can be derived as
\begin{equation}
    d_2 = \frac{(L_{amb2}-L_{m2}^{'})L_{amb1} + L_{m1}^{'}L_{amb2}}{L_{amb2} - L_{amb1}}
\end{equation}

Since $\Delta f_{rep}$ of microcombs generated in resonators with the same size usually falls in the range of several MHz, this method can increase the measurement distance by three orders of magnitude.

A prominent requirement of this method for detectors is the bandwidth must cover of the beat signal between the two microcombs. The upper bound of the frequency is approximately 1/2 FSR of the combs. According to Nyquist's law, the sampling rate of an oscilloscope needs to be greater than twice the highest frequency of the RF-comb output by the photodetector. Typically, the FSR of frequency combs based on micro-resonators ranges from tens to hundreds of GHz, which is difficult to achieve. To address this issue, we use microcombs generated by a shared laser, referred as coherent microcombs, to reduce the frequency of the RF-comb. To avoid overlap of positive and negative frequencies, an appropriate frequency $F$, which is covered by low-bandwidth detectors, is chosen as the difference between pump lasers of two microcombs. Therefore, the beat signal is an RF-comb centered on $F$ with spacing $\Delta f_{rep}$. As mentioned before $\Delta f_{rep}$ usually falls in the range of several MHz. Typically, the range of the RF-comb can be limited in several GHz, which possibly meet the bandwidth of FPGAs to realize real-time processing. Furthermore, it is possible to adjust $F$ according to the value of $\Delta f_{rep}$ to meet the minimum requirements for the detectors. 

\section{Coherent dual-comb generation}

\begin{figure*}[t]
    \centering
    \includegraphics[width=\textwidth]{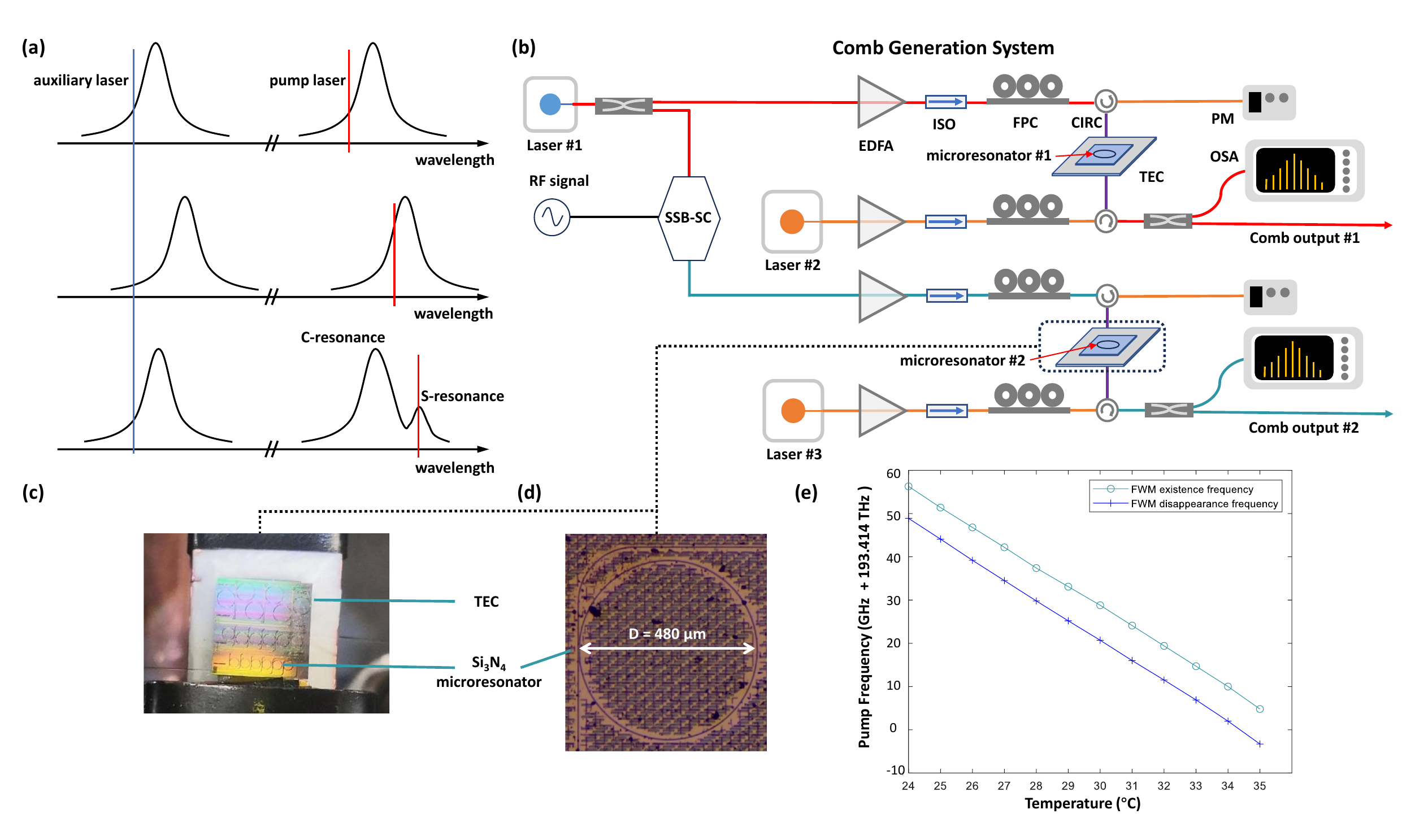}
    \caption{(a) Soliton generation process in dual-pump method. The auxiliary laser lies in the blue detuned regime while the pump laser scan across the resonant peak. The single soliton causes a deformation of the resonant peak, which consists of a C-resonance and a S-resonance, and the pump frequency lies in the S-resonance. (b) Experimental setup. TSL, tunable semiconductor laser; EDFA, Erbium-droped fiber amplifier; Iso, isolator; FPC, fiber polarization controller; TEC, thermoelectric cooler; PM, powermeter; SSB-SC, suppressed carrier single-sideband modulator; OSA, optical spectrum analyzer. (c) Photograph of the $Si_3N_4$ chip placed on a TEC. The waveguide of the chip is coupled with a lensed fiber. (d) Microscopy image of a microresonator with the diameter of 480 $\mu$m. (e) Resonant frequency as a function of temperature when changing from 24°C to 35°C.}
    \label{OFC}
\end{figure*}

Dual-pump method\cite{xjzz-2019-8-437} is applied in our soliton generation system. Each microresonator is pumped by a pump laser and an auxiliary laser. The pump laser is used to generate the microcomb, and the auxiliary laser is used to balance the thermal dynamics during soliton generation as well as precisely control the detuning. The auxiliary laser lies within the blue detuned regime and possesses self-thermal locking, which enables high precise control of pump detuning \cite{Geng:20}. The principle is shown in Fig.\ref{OFC}(a)\cite{Zhang:19}. As the auxiliary laser enters the resonance peak from blue-detuned regime, the resonator is heated with redshift of resonant frequency. During the generation of soliton microcombs, the wavelength of auxiliary laser is set as a proper wavelength and power to balance the heat flow caused by the pump laser. Consequently, the pump laser scan across the resonance peak and enter the soliton existence region in red-detuned range with little thermal variation, and a DKS is generated in the microresonator.

An additional demand for generating coherent microcombs is to simultaneously match the pump frequency with the resonance of two resonators. Previous research uses acoustic optical modulator (AOM) to control the frequency difference between pump lasers\cite{Geng:22}. In this work, thermo-optic tuning to microresonators, with changing of effective cavity length, is a convenient route to generate coherent dual microcombs. Limited by the manufacturing process, even resonators processed according to the same structure are shown to have independent resonance. However, with temperature control to microresonators, the resonance can be precisely adjusted to match the pump laser. Fig.\ref{OFC}(b) illustrates the experimental setup for dual soliton microcombs generation system, which is an upgraded version of a standalone soliton microcomb prototype based on dual-pump method \cite{Xi:22}. There are two microresonators (mark as 1\# and 2\#) and three 1550-nm narrow linewidth lasers (mark as 1\#, 2\# and 3\#) in the system. Laser 1\# works as the pump laser to generate two microcombs. It is split into two parts and one part is directly sent to an Er-doped fiber amplifier (EDFA) while the other one is sent to a suppressed carrier single-sideband (SSB-SC) modulator. A single-sideband is generated with an adjustable frequency difference and sent to an EDFA. Both parts are boosted to an appropriate power and pump the microresonators. Laser 2\# and 3\# work as the auxiliary lasers. They are amplified and injected into the microresonators from the opposite direction of the pump laser. The FPCs are used to control the polarization, which is necessary for single soliton microcomb generation. We adopt two $Si_3N_4$ micro-resonators on two chips for soliton microcombs generation. The cross sections of each chip are about $5\,mm \times 5\,mm$. The diameter of the microresonator is 240 $\mu m$ and FSR is approximately 95 GHz. To support DKS formation, the resonators are designed to have anomalous dispersion and high Q-factor. Each $Si_3N_4$ chip is coupled with two lensed fiber and placed on a thermoelectric cooler (TEC), which is used to enhance the stability as well as tune $f_{rep}$ and $f_{\mu=0}$, which stands for the frequency of the tooth corresponding to the pump laser. As the TEC temperature raised, the microresonator is lengthened and the repetition rate and the resonant frequency are increased. It is worth mentioning that the frequency difference between pump lasers is only limited by the RF signal generator and the bandwidth of SSB-SC modulator. In a specific ranging system, it would be reset according to actual requirements.

Before generating coherent microcombs, we roughly measure the redshift trend of the resonant frequency at high power and the variable range of $f_{rep}$ with the influence of thermal optical effect. With 100 MHz each step, a laser scan across the resonant peak. The existence of first pair of four wave mixing sidebands is regarded as the frequency entering the resonant peak, while the frequency of leaving the resonant peak is presented by the disappearance of the chaotic spectrum. Due to the slow tuning speed and the influence of thermal effects, the intracavity light field does not enter the soliton state. As shown in Fig.\ref{OFC}(e), the red-shift trend of the resonance peak shows a roughly linear trend with different temperature. For every 1°C increasing in temperature, the red shift of the resonance peak is approximately 4.8GHz. Therefore, for the 95GHz microresonators used in this experiment, a simple and feasible solution to tune the resonant peak to any frequency is changing the temperature within 20°C.

In order to measure the variation of repetition rate as temperature changes, the interferometric method is utilized.  Two free-running single soliton microcombs are generated through dual-pump method. As is well-known that the repetition rate is possibly influenced by the intracavity power, the pump power remains unchanged during the whole process. For one microcomb, power of the pump laser is set as 27.5 dBm, and the auxiliary laser is set as 30.5 dBm. For the other, the pump and auxiliary laser are set as 29 dBm and 32.5 dBm, respectively. The TECs are both tuned within the range of 25°C to 40°C. Since each one of the single solitons possibly disappears when detuning changes with the temperature, the lasers must be tuned again after the stabilization of temperature. Fortunately, the automated soliton generation system is available \cite{computer-controlled}. One TEC temperature is changed at a time while the other one stays constant. It is found that the tunable range of the repetition rate can reach several MHz. When the TECs tuned from 25°C and 35°C to 30°C and 36.5°C, the repetition rate difference between the two free-running microcombs varies from 1.84 MHz to 14.58 MHz, which shows a variable range of at least 12 MHz. The variation is not continuously observable because the bandwidth of our detectors is 18 GHz while the signal may reach 50 GHz. Though larger range of temperature is available, the repetition rate difference does not change significantly, which may because of the slow heat conduction of our chip. Herein, we would further extend the tunable range by changing the temperature as well as the pump power.

\begin{figure*}[t]
    \centering
    \includegraphics[width=\textwidth]{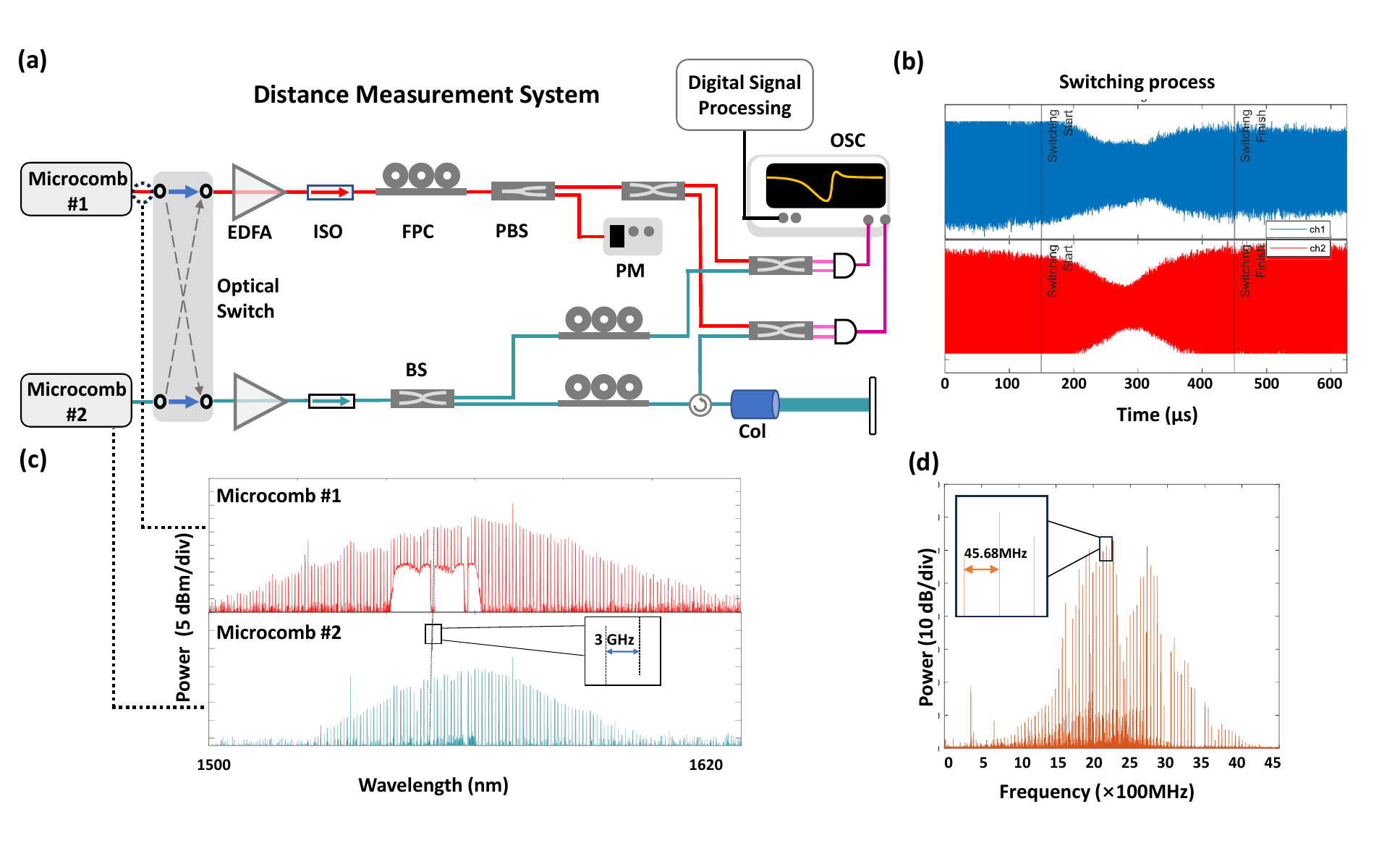}
    \caption{Experimental demonstration and recorded signal. (a) Experimental setup. DKS microcombs are generated and filtered by FBGs to reflect the pump and auxiliary laser. Then, they are sent to a $2 \times 2$ optical switch and the ranging system. Col, collimator. PBS, polarization beam spilter. (b) The entire recorded signal in 625 $\mu s$, including 3 parts: two series of periodic pulses (first and last 150 $\mu s$) and a switching process (approximately from 150 $\mu s$ to 450 $\mu s$) (c) Coherent microcombs used in the experiment, with 3 GHz center frequency difference. (d) The Fourier transform of the measurement path signal (ch1) (only 0-4.5 GHz shown). The spacing between the adjacent teeth is 45.68 MHz.}
    \label{Experiment-results}
\end{figure*}

Here, we know the ability of the thermal optical effect to adjust the parameters of the optical frequency comb. Coherent microcombs can be generated through the following steps. First, input a 3 GHz RF signal to the SSB-SC modulator to set the pump frequency difference as 3 GHz. Then, adjust the temperature to make the resonant frequency difference between the two microcavities roughly equals to the pump laser frequency difference (3 GHz in out experiment). Third, the two auxiliary lasers are tuned to the blue detuning region, generating a small number of sidebands through four-wave mixing. One pump laser scans through the resonant peak, generating a single soliton in one of the micro-resonators. For the other microresonator, the auxiliary laser is used to tune the detuning of the pump laser, making the pump laser reach the single soliton state. Specifically, when the microresonator \#1 reaches a single soliton state, if the optical field inside the microresonator \#2 is in a chaotic or multi-soliton state, the auxiliary laser wavelength should be decreased to make the resonance blue-shifted, increasing the pump laser detuning, and causing the intracavity optical field to enter single soliton state. If the pump laser has completely passed through the resonant peak of the microresonator \#2, it is necessary to first increase the auxiliary laser wavelength, causing the resonance red-shifted until the pump laser re-enters the resonance. At this point, the intracavity optical field may be in chaotic state or multi-soliton state. Then, decreasing the auxiliary laser wavelength to increase the detuning of the pump laser and enters single soliton state.

In this way, we obtained two optical frequency combs with a center frequency difference of 3 GHz. With temperature control the microcombs keep stable for several hours, which fully meet the requirements of the ranging system. Due to the limited nonlinear conversion efficiency of dissipative Kerr solitons, the soliton state optical frequency comb has a strong continuous wave background at the pump frequency, containing most of the power of the optical frequency \cite{Conversion-efficiency}. Due to the smoothness of spectral envelopes is required in multi-longitudinal-mode heterodyne interferometry, the pump light and the reflected light of the auxiliary laser are filtered out through fiber Bragg gratings (not shown). Consequently, the coherent microcombs become an ideal light source for dual-comb ranging.

\section{Dual-comb ranging}

\begin{figure*}[t]
    \centering
    \includegraphics[width=\textwidth]{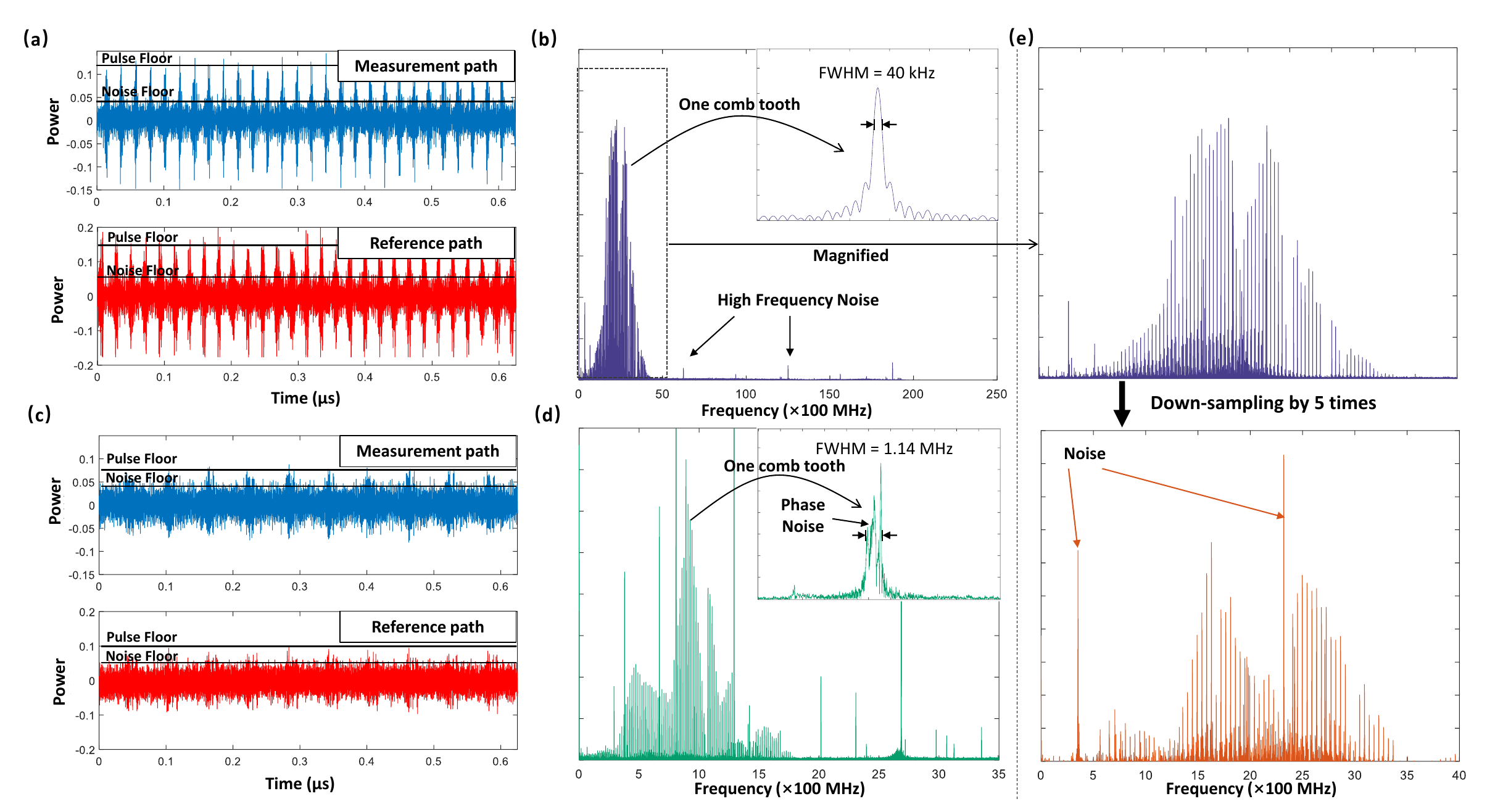}
    \caption{Comparison between coherent dual-comb raging and free-running dual-comb ranging. (a) 0.625 $\mu s$ of recorded signals when using two coherent microcombs. The signal-to-noise rate is about 4 in both paths. (b) The Fourier transform of signal in (a). The FWHM is 40 kHz. (c) 0.625 $\mu s$ of recorded signals when using two free running microcombs. The signal-to-noise rate is lower than 2 in both paths. (d) The Fourier transform of signal in the measurement path in (c) (only 0 - 2 GHz shown). The full width at half maximum (FWHM) is 1.14 MHz. (e) Comparison between the Fourier transform of the original data (with sampling rate at 50 GSa/s) and the data down-sampled by 5 times. Several high-frequency noise with higher frequencies in the RF spectrum appears at lower frequencies after down-sampling.}
    \label{comparison}
\end{figure*}

The multiheterodyne detection devices for distance ranging are shown in Fig.\ref{Experiment-results}(a). Microcomb 1\# and 2\# are the LO comb and signal comb in the system, respectively. The signal comb and the local oscillator (LO) comb are amplified to about 6 mW by a pair of C+L-band EDFAs. The signal comb and LO combs are split by two fiber couplers. The polarization beam spilter (PBS) and PM are used to separate a linear polarization mode to participate in the interference. In the measurement path, one part of the signal comb is routed to the target and back, then forwarded to a BPD after interfering with the LO comb. In the reference path, the other part of signal and LO combs directly interferes with each other and sent to another BPD, while the FPCs guarantee the premise of interference.

For a precision analysis, we measure the distance between the collimator and a static mirror. Fig.\ref{Experiment-results}(c) shows the spectrum of two soliton microcombs that are pumped by laser with 3 GHz frequency difference. The pump and auxiliary laser are filtered by fiber Bragg gratings, so the relative spectral lines are missing. Since different proportion of power are coupled into the OSA, the recorded spectrum looks different. Fig.\ref{Experiment-results}(b) shows the signal recorded by the oscillator. The sample rate of the oscillator is 50 GSa/s and entire recording time is 625 $\mu s$ (31250000 samples in total), including the switching time (about 300 $\mu s$). Notice that the switching process must be reversible to meet the needs of practical applications. The signal from both paths are presented as pulse series, and the first 0.625 $\mu s$ are magnified in Fig. \ref{comparison}(a), which can be regarded as the sampling of the signal comb by the LO comb.

Fig.\ref{Experiment-results}(d) shows the Fourier transform of the beat signal recorded in the measurement path between the signal and LO combs, which is an RF comb in the frequency domain. As mentioned before, the distance information is contained in the phase difference between RF combs transformed by the two paths. The spacing of beat notes equals the repetition rate difference and amounts to $\Delta f_{rep}=45.68\,MHz$. Thereby, the minimum acquisition time is $t_{acq}=1/\Delta f_{rep}=22.22 \,ns$. Limited by the switching time, the acquisition frequency is about 3.3 kHz and the ambiguity distance is $L_{amb} = c/2\Delta f_{rep} = 3.28\, m$. Though the extension of ambiguity range sacrifice the acquisition rate, the switching process contains enough time to finish the digital signal processing, which provides the potential to real-time processing. At an average time of 9.56 $\mu$s, the Allan standard deviation is $\sigma = 346\,nm$. The uncertainty is $\delta = \sigma / L_{amb} = 346\,nm/3.28\, m \approx 1.05 \times 10^{-7} $. Additionally, the more long-range distance measurement with high precision can be achieved by combing our scheme and other ranging method, such as chaotic dual-comb ranging\cite{lukashchuk2023chaotic}.

To better illustrate the advantages of using coherent microcombs in the dual-comb ranging systems comparing to using free-running microcombs. We analyze the ranging signal in both the time and the frequency domain. The phase noise significantly decreased, because the pump lasers of coherent microcombs are generated by the same laser, while one of them is amplified after carrier-suppression single-sideband modulation. For the time domain, when coherent microcombs are used, Fig.\ref{comparison}(a) and (c) shows the recorded signal in both paths when the combs are pumped together and separately, respectively. The ratio of pulse amplitude to noise amplitude is twice as it when using free-running microcombs. Therefore, signal-to-noise ratio has increased by two times. As shown in Fig.\ref{comparison}(b) and (d), in the frequency domain, the 3 dB bandwidth of a single spectral line of the RF-comb is approximately 1.14 MHz when pumped separately, which contains lots of noise. Under the same pump, the 3 dB bandwidth of a single spectral line is about 40 kHz, which decreases by about 29 times.

The difference between the center frequencies of coherent microcombs is  3 GHz. The RF comb generated in this way is below approximately 4 GHz, and the corresponding oscilloscope sampling rate only needs 8 GHz, which greatly reduces the dependence on the high sampling rate of the oscilloscope and the high bandwidth of the photodetector. To verify the feasibility of the scheme, we downsampled the data obtained from the ranging experiment to 10 GSa/s before processing, and obtained the same results as before. The RF spectrum after downsampling is shown in Fig.\ref{comparison}(e), showing the complete spectrum information is preserved. Since no pre-anti-aliasing filter is used, aliasing makes several high-frequency noise appearing at low frequencies in the RF spectrum after being downsampled, but these lines would not affect our data processing. The advantage of this method lies in greatly reducing the dependence on sampling rates of oscilloscopes and bandwidth of photodetectors, and making it possible for data processing to run on common field-programmable gate
arrays (FPGA) integrated with lower sampling rate analog-to-digital conversion modules (ADCs) or specialized hardware.

\section{Conclusion}

We demonstrate the viability of coherent soliton microcombs to act as the optical source for phase-based optical ranging system and a key step toward real-time dual-comb ranging. Although most of the technical blocks of integrated dual-comb ranging are demonstrated, one of the remaining key challenge is achieving real-time processing. We have solved this problem in terms of reducing the amount of data required for distance measurement. While maintaining the accuracy of distance measurement and the acquisition rate, we have achieved an improvement in processing speed by lowering the frequency of RF signals. Alternatively, our scheme is a modularized system, including the comb-generation module and distance-measurement module, which makes it reconstructible. Other pumping method, such as the automated self-injected lock method\cite{self-injection-locking}, the laser-cavity frequency comb with high power efficiency\cite{bao:hal-02417145} can be used, permitting high quality optical frequency combs. Based on these studies, we believe that the real-time dual-comb LIDAR could have further application fields. Real-time signal processing also enables real-time results acquisition for dual-comb based imaging or identification.

\begin{acknowledgments}
This work is supported by the Key-Area Research and Development Program of Guangdong Province under Grant 2018B030325002, the National Natural Science Foundation of China (Grant No.62075129), the Open Project Program of SJTU-Pinghu Institute of Intelligent Optoelectronics (No.2022SPIOE204) and the Science and Technology on Metrology and Calibration Laboratory (Grant No.  JLJK2022001B002).
\end{acknowledgments}



%

\end{document}